# Emulating exceptional-point encirclements using imperfect (leaky) photonic components : Asymmetric mode switching and omni-polarizer action

J. B. Khurgin,[1] Y. Sebbag,[2] E. Edrei,[2] R. Zektzer,[2] K. Shastri,[3] U. Levy,[2] and F. Monticone[3,*]

[1] *Department of Electrical and Computer Engineering, Johns Hopkins University, Baltimore MD 21218, USA*

[2] *Department of Applied Physics, The Hebrew University of Jerusalem, Jerusalem, 91904, Israel*

[3] *School of Electrical and Computer Engineering, Cornell University, Ithaca, New York 14853, USA*

* francesco.monticone@cornell.edu

**Abstract –** Non-Hermitian systems have recently attracted significant attention in photonics. One of the hallmarks of these systems is the possibility of realizing asymmetric mode switching and omni-polarizer action through the dynamic encirclement of exceptional points (EP). Here, we offer a new perspective on the operating principle of these devices, and we theoretically and experimentally show that linear asymmetric mode switching and omni-polarizer action can be easily realized – with the same performance and limitations – using simple configurations that emulate the physics involved in encircling EP's without the complexity of actual encirclement schemes. The proposed concept of "encirclement emulators" and our theoretical and experimental results may allow a better assessment of the limitations, practical potential, and applications of EP encirclements in non-Hermitian photonics.

## 1. INTRODUCTION

Over the past two decades, the concept of parity-time (PT) symmetry has been at the center of attention in different areas of science and engineering, motivated by the realization that a system described by a non-Hermitian Hamiltonian can have real eigenvalues, provided that the Hamiltonian satisfies the PT-symmetry conditions [1]. The entire new discipline of non-Hermitian quantum mechanics has since been developed [2, 3], with a particular focus on so-called exceptional points (EPs) [4], i.e., degeneracy points where two or more eigenvalues *and* eigenfunctions of the Hamiltonian coalesce. While efforts on extending quantum mechanics to the complex domain have been hampered by the lack of experimental observation of PT-symmetric properties in nature, these properties turned out to be relatively easy to "synthesize" in classical wave-physics systems, especially in photonics, where non-Hermiticity arises from the presence of optical gain and/or loss, corresponding to the imaginary part of the dielectric constant (or radiation loss, parametric gain, etc.). Following the first pioneering theoretical works in this field [5-8], experimental observations of PT-symmetric phenomena in photonics, including EP's, were soon reported [9, 10]. Since then, PT symmetry has been demonstrated in large variety of photonic structures, including coupled waveguides, coupled resonators, lasers [11], Bragg gratings [12] etc. (more information can be found in extensive reviews [13-15]. In particular, the peculiar behavior of a system near EPs [16] has been observed and discussed in several papers. Notably, it has been shown that, in the vicinity of an EP, the sensitivity of various sensors [17-19] can be enhanced due to the square-root branch-point nature of these degeneracies, suggesting that PT symmetry and EPs are more than just a fundamental physical concept, but may potentially be useful for relevant applications.

One of the most fascinating properties of PT symmetric systems is the behavior of the eigenstates upon the encirclement of an EP through a variation of the system's parameters in space or time. Adiabatic encirclement of EPs was first studied in [20, 21] and it was predicted that, each time an adiabatic encirclement takes place, it will cause a flip of the eigenstates. This state-flipping behavior has been confirmed in a number of "quasi-static" experiments where the eigenstates were monitored, under stationary conditions, while discretely changing the parameters of the system [22-24]. However, in recent years, it has been shown theoretically that, when the encirclement occurs dynamically, the fact that the Hamiltonian of the system is non-Hermitian (i.e., in optics it involves loss and gain) prevents one from applying the adiabatic theorem, and the encirclement of the EP results in a different final state that depends only on the direction (helicity) of the encirclement and not on the initial state [25-28]. This behavior suggests intriguing implications for multi-port or multi-mode devices whose output is independent of the input state. The dynamic encirclement was first experimentally demonstrated in 2016 in non-Hermitian (lossy) microwave waveguides with coupled modes, resulting in asymmetric mode switching [29]. Very recently this concept of dynamic EP encirclement was also proposed to realize a linear asymmetric omni-polarizer, in which arbitrarily polarized input light always exits the system in the same polarization state going in one direction and in the orthogonal state going in the opposite direction. This idea was originally proposed in [28] and experimentally demonstrated in [30]. While omni-polarizers have already been designed and demonstrated based on *nonlinear* effects, for example using optical fiber nonlinearities in [31], the recent works in Refs. [28],[30], as well as this Letter, are instead focused on the possibility of realizing *linear* asymmetric omni-polarizers using EP encirclements or some other schemes. These first results on asymmetric mode switching and omni-polarizer action have understandably generated a certain amount of excitement for the development of robust linear photonic devices based on dynamic EP encirclement. We believe it is therefore important to assess the practicality of these devices as well as their uniqueness, i.e., to answer the question about whether the desired functionality can only be achieved by dynamic EP encirclement, and not with simpler photonic schemes. In this Letter, we answer these relevant questions based on general theoretical considerations and proof-of-concept experiments. We show that linear asymmetric-dynamic-encirclement devices suffer from a major inherent drawback: while their output state is indeed mostly independent of their input state, their transmission amplitude necessarily exhibits a very strong dependence on the input state. Furthermore, we demonstrate experimentally that asymmetric mode-switching performance virtually identical to dynamic encirclement schemes can be achieved in far simpler optical systems using few standard and ubiquitous linear components like couplers, beam-splitters, and waveplates, and whose operational principle can be understood without involving exceptional points.

## 2. RECIPROCITY AND TRANSMISSION-MATRIX ASYMMETRY

The recently proposed and experimentally demonstrated asymmetric mode-switches [29] and omni-polarizers [28, 30] are linear reciprocal devices that produce a certain output state (A), in the forward direction, regardless of the input state (A or B), whereas they produce the opposite output state (B) in the backward direction, again independent of the input state. The state (A) or (B) can be a certain waveguide mode, polarization state, or any other state belonging to a set of orthogonal channels, e.g., a state of angular momentum.

The transmission properties of an asymmetric 2-port linear device, as in Fig. 1(a), can be described by a 2x2 transmission matrix (composed of the transmission elements of the scattering matrix)

$$T^{\rightarrow} = \begin{pmatrix} T^{\rightarrow}_{AA} & T^{\rightarrow}_{BA} \\ T^{\rightarrow}_{AB} & T^{\rightarrow}_{BB} \end{pmatrix}. \qquad (1)$$

The device may contain gain and loss, hence there is no passivity constrains on the magnitudes of the transmission coefficients, i.e. the scattering matrix is not unitary; however, the device is assumed to be reciprocal, hence the transmission matrix in the backward direction is

$$T^{\leftarrow} = \begin{pmatrix} T^{\leftarrow}_{AA} & T^{\leftarrow}_{BA} \\ T^{\leftarrow}_{AB} & T^{\leftarrow}_{BB} \end{pmatrix} = \begin{pmatrix} T^{\rightarrow}_{AA} & T^{\rightarrow}_{AB} \\ T^{\rightarrow}_{BA} & T^{\rightarrow}_{BB} \end{pmatrix}. \qquad (2)$$

Consider a device designed to produce output A when excited from the left and output B when excited from the right. Reciprocity (symmetry of the scattering matrix) imposes severe constraints on the operation of such a device. In particular, an ideal *asymmetric* mode-switch, with transmission amplitude in a certain output state *equal* for both input states, for both excitation directions, must be nonreciprocal. To understand this better, let us assume unity transmission coefficient from port B on the left to port A on the right, i.e., $T^{\rightarrow}_{BA} = 1$; then, it is necessary that $T^{\rightarrow}_{AA} = \alpha$, with $\alpha \ll 1$ if the device is to produce (mostly) output B in the opposite direction. Indeed, because of reciprocity, $T^{\leftarrow}_{AB} = T^{\rightarrow}_{BA} = 1$ and $T^{\leftarrow}_{AA} = T^{\rightarrow}_{AA} = \alpha$. Therefore, if $\alpha = 1$, the device can operate as an ideal mode-switch or omni-polarizer in the forward direction, with identical transmission amplitude for both input states (A or B), but it would then act as a signal splitter in the backward direction. To put it more bluntly, an ideal mode switch or omni-polarizer must be described by a forward transmission matrix

$$T^{\rightarrow} = \begin{pmatrix} 1 & 1 \\ 0 & 0 \end{pmatrix}, \qquad (3)$$

which, however, would mean that, since the device is reciprocal, the backward transmission matrix would be

$$T^{\leftarrow} = \begin{pmatrix} 1 & 0 \\ 1 & 0 \end{pmatrix}, \qquad (4)$$

which is not an omni-polarizer or mode switch, but a mode filter followed by a 3dB mode splitter, i.e., something of dubious utility. Indeed, experimental results in [29] show that $\vec{T}_{BA}/\vec{T}_{AA} = 1/\alpha = 28dB$ and $\vec{T}_{AB}/\vec{T}_{BB} = 14dB$. Assuming that one can bring $\vec{T}_{BA}$ to unity using optical amplifiers, the fact that insertion loss for one of the input states is as high as nearly 30dB puts practicality of such a device in doubt.

Thus, any linear reciprocal device designed to operate as an *asymmetric* mode-switch or *asymmetric* omni-polarizer must have transmission amplitude that is strongly dependent on the input state. This is true regardless of the device implementation, and it follows directly from reciprocity. In light of these considerations, we argue that, if only reciprocal linear components are used, there are much simpler ways to obtain the same asymmetric mode-switching action than dynamic encirclement of EP's.

## 3. EXCEPTIONAL-POINT ENCIRCLEMENTS AND THEIR EMULATION

Let us now try to understand the basic physical principle that causes asymmetric mode switching during an EP encirclement, and whether this process can be emulated. For two coupled modes $E_{1,2}$ with complex propagation constants $\beta_{1,2} + i\gamma_{1,2}$, the basic, yet rather prolific, characteristic matrix (Hamiltonian) that is essentially at the basis of the entire PT-symmetry field is

$$H = \bar{H} + H_{int} = \begin{pmatrix} \bar{\beta} + i\bar{\gamma} & 0 \\ 0 & \bar{\beta} + i\bar{\gamma} \end{pmatrix} + \begin{pmatrix} \delta\beta + i\delta\gamma & \kappa \\ \kappa & -\delta\beta - i\delta\gamma \end{pmatrix}, \quad (5)$$

where $\kappa$ is the coupling coefficient, $\bar{\beta} = (\beta_1 + \beta_2)/2$, $\bar{\gamma} = (\gamma_1 + \gamma_2)/2$, $\delta\beta = (\beta_1 - \beta_2)/2$, $\delta = (\gamma_1 - \gamma_2)/2$. The two complex eigensolutions, $\tilde{\beta}_\pm = \bar{\beta} + i\bar{\gamma} \pm \sqrt{(\delta\beta + i\delta\gamma)^2 + \kappa^2}$, reveal two EP's (shown in Fig. 1(b)) for $\delta\beta = 0$ and $\kappa = \pm\delta\gamma$, at which two solutions have identical eigenvalues and eigenvectors.

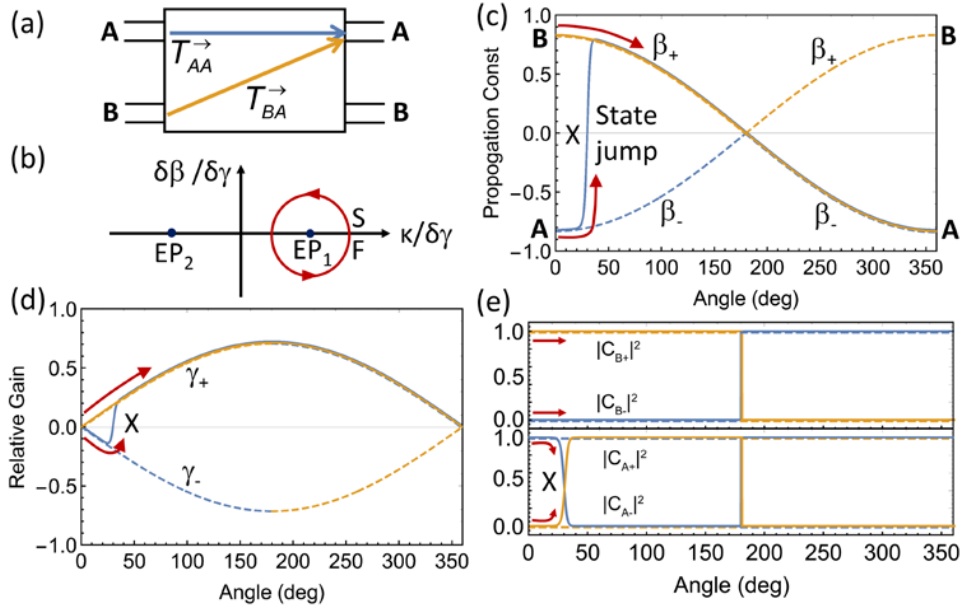

Figure 1. (a) 2-port linear system under study: ports A and B may correspond to different guided modes or polarization states. (b) EP's and encirclement trajectory. Evolution during encirclement of (c) relative propagation constants, (d) relative extinction coefficient (dashed curves: eigensolutions; solid curves: eigensolutions weighted by the respective population coefficient), and (e) population coefficients for the system initially prepared in mode B (top) and mode A (bottom). Red arrows indicate the direction of encirclement.

Let us then consider what happens when we encircle one of the EP's, counterclockwise from a starting point S to a final point F = S. The results are shown in Figs. 1(c)-(e) for an encirclement along the contour with radius $0.3\delta\gamma$. In Figs. 1(c),(d) one can see the evolution of the relative propagation constant, $\beta_\pm = \text{Re}(\tilde{\beta}_\pm) - \overline{\beta}$, and the relative gain, $\gamma_\pm = \text{Im}(\tilde{\beta}_\pm) - \overline{\gamma}$, while Fig. 1(e) shows the evolution of the population coefficients $|C_{\pm A,B}|^2$, i.e., the projection of the instantaneous state vector along the instantaneous eigenvectors $E_\pm(\theta)$. If we initially prepare the system in one of the eigenstates, $E_-(0)$ (or $E_+(0)$), associated with mode A(B), and assuming full adiabaticity, the mode should be converted into mode B(A) as a result of a full parametric encirclement. However, as the two instantaneous eigenstates evolve, they experience very different relative gain/loss; in addition, due to the non-Hermitian nature of the problem the eigenstates are not strictly orthogonal, and a portion of energy may leak from one state to the other. In particular, a leakage from the lossy state (originating from A, see Fig. 1) to the amplifying state cannot be ignored, as this fraction of energy will then experience a relative gain, while the remaining fraction in the lossy state will suffer extinction. This induces a state jump around point X in Fig. 1, consistent with Ref. [28], [29] (the location of point X depends on the degree of non-adiabaticity, i.e., the speed of dynamic encirclement). As a result, toward the end of a full dynamic encirclement, the optical signal originating from A will be much stronger in mode A than in mode B. Conversely, the light that leaked from the amplifying state, originating from B, into the lossy state would suffer extinction and can be neglected (no state jump occurs in this case). Hence, *no matter where the signal originates from (i.e., its input state), it ends up mostly in mode* A, as indicated by the solid curves in Figs. 1(c-e). Following the same reasoning, it is also clear that reversing the direction of encirclement (or of propagation) will bring the signal mostly into mode B, regardless of the input state. This is the mechanism behind the asymmetric mode-switching effect in [29]. However, it is also important to realize that, when dynamic encirclement takes place, along one trajectory (going from B to A) the entire signal propagates with relative gain, while along the other possible trajectory with the same output (going from A to A) at first the signal experiences large loss (attenuation and leakage) and then it gets recovered by experiencing relative amplification. This combination of high loss followed by high gain may look benign to a theorist, yet it appears noxious in practice since the signal to noise ratio (SNR) is greatly deteriorated.

At this point, an astute reader may have noted that since all the "modal switching" action takes place in the vicinity of point X, the relative gain/loss preceding and following it does not have to be distributed and can simply be formed by lumped optical amplifiers and attenuators. Theoretically speaking, in the absence of gain/loss, adiabaticity for slowly varying systems tends to get restored, but in any real system ideal adiabaticity is never fully achieved. It is precisely this fact (leakage due to imperfect adiabaticity) that makes the remarkable properties of dynamic encirclement possible, while the specific evolution of the eigenmodes before/after point X is insignificant, as long as different trajectories have different gain and loss. Based on these considerations, it is clear that essentially the same behavior can be observed and emulated by using a leaky mode-coupler preceded and followed by lumped gain/loss elements. As illustrated in Fig. 2(a), the coupler can simply be an imperfect directional coupler with coupling coefficient $|\kappa|<1$, preceded and followed by optical amplifiers and optical attenuators (using two different levels of loss would also work). The transmission matrices for the proposed "encirclement emulator" are

$$T^{\rightarrow} = \begin{pmatrix} 1-\kappa^2-\gamma_1+\gamma_4 & \kappa^2+\gamma_3+\gamma_4 \\ \kappa^2-\gamma_1-\gamma_2 & 1-\kappa^2-\gamma_2+\gamma_3 \end{pmatrix}, \quad T^{\leftarrow} = \begin{pmatrix} 1-\kappa^2-\gamma_1+\gamma_4 & \kappa^2-\gamma_1-\gamma_2 \\ \kappa^2+\gamma_3+\gamma_4 & 1-\kappa^2-\gamma_2+\gamma_3 \end{pmatrix}, \quad (6)$$

for forward and backward propagation, respectively. The configuration in Fig. 2(a) offers a wide range of flexibility for obtaining the desired coupling ratios. For our experimental demonstration of such an "encirclement emulator", we have chosen to use no amplifiers, i.e., $\gamma_3 = \gamma_4 = 0$, a 3dB optical coupler, $\kappa^2 = 0.5$, and attenuators with $\gamma_1 = 25dB$ and $\gamma_2 = 15dB$, as shown in the illustration of our experimental setup in Fig. 2(b). The transmission matrix of the system is therefore

$$T^{\rightarrow} = \begin{pmatrix} -28dB & -3dB \\ -43dB & -18dB \end{pmatrix}, \quad T^{\leftarrow} = \begin{pmatrix} -28dB & -43dB \\ -3dB & -18dB \end{pmatrix}, \quad (7)$$

which clearly indicates that for forward (backward) propagation the output is mostly in mode A(B). The experiment was performed using a standard 3dB fiber coupler and two variable optical attenuators (VOA's) in ports A-left and B-right, but of course, an integrated version of the device can be easily envisioned. . As illustrated in Fig. 2(b), a tunable laser source (New-port, Velocity TLB-6700) was injected separately through each port while the intensity was measured at both outputs using a standard silicon detector (Thorlabs, DET100A). As the dynamic range of the detector is limited, pre-calibrated external filters were introduced and accounted for post-processing when needed. To eliminate laser intensity fluctuations, a reference arm was used to normalize each measurement.

The experimental results are reported in Fig. 2(c), where all data points are presented relative to perfect transmission, i.e., without attenuators and coupling loss. If one compares these measurements with the observations of dynamic encirclement reported in [29], and replicated in Fig. 2(c), the results are identical (except for the fact that our common insertion loss is more than 20 dB smaller) : almost all forward (backward) propagating light ends up in mode A(B). Notice that the transmission amplitude is highly asymmetric both in our scheme and in Ref [29], with $T_{AA}^{\rightarrow}/T_{BA}^{\rightarrow} = \alpha \approx -25dB$, and $T_{BB}^{\leftarrow}/T_{AB}^{\leftarrow} \approx -15dB$. As discussed above, this asymmetry is *unavoidable* in any linear and reciprocal asymmetric mode-switch. The asymmetry can be mitigated by using nonlinear saturable amplifiers in positions $\gamma_3, \gamma_4$ which simply add a certain power to the signal no matter what the input is. Then, a more uniform transmission amplitude for different inputs can be attained; however, this strategy would not alleviate the main impediment to practical applications, namely, the fact that the signal following one of the pathways initially suffers a significant loss that deteriorates the SNR.

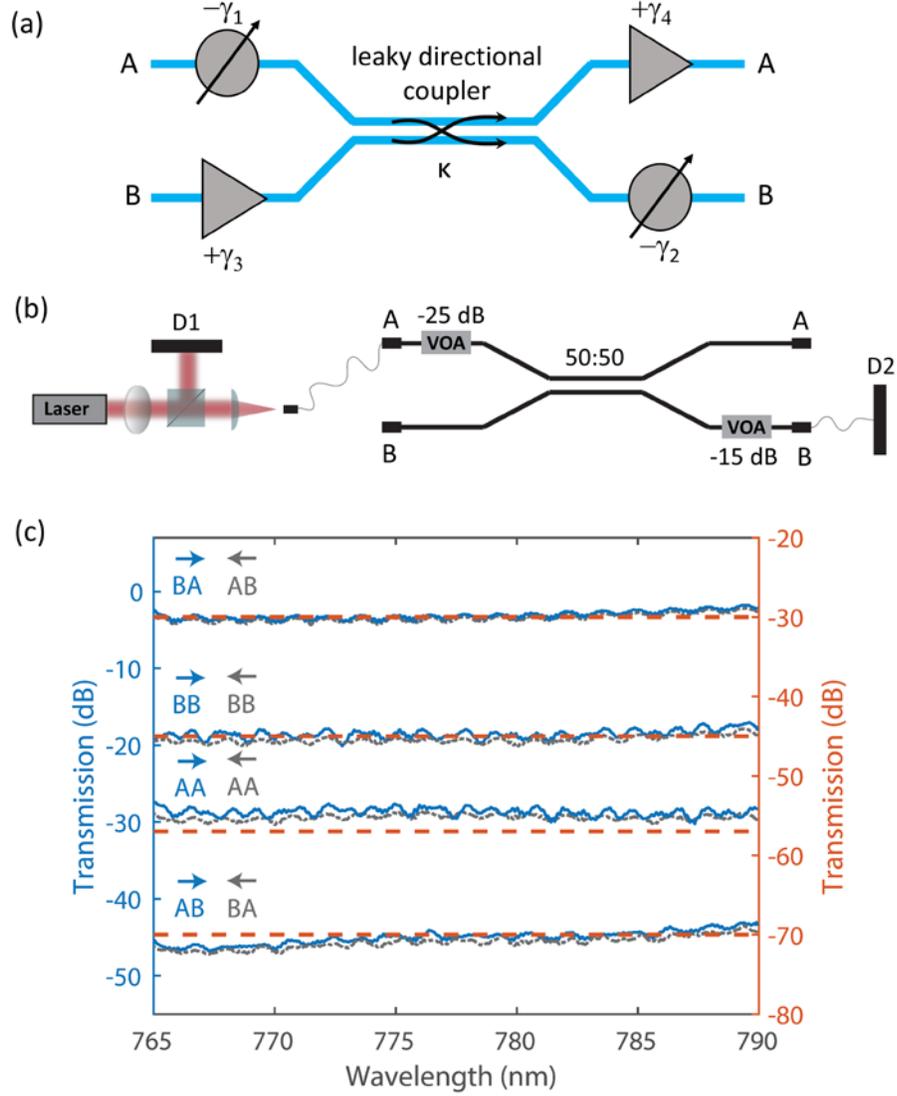

Figure 2. (a) Proposed scheme realizing asymmetric mode switching, emulating a dynamic EP encirclement. (b) Experimental setup described in the main text. (c) Experimentally measured transmission amplitudes (blue) for a configuration as in (b), compared to the transmission amplitudes for the device in [29] (orange; values taken at the central wavelength).

## 4. ASYMMETRIC OMNI-POLARIZER

The approach described in the previous section to emulate the effects of an EP encirclement can be easily applied to other tasks, such as to realize the linear asymmetric omni-polarizers first explored in Ref. [28] and recently realized in [30]. A possible configuration is shown in the dashed box in Fig. 3(a), based on two polarization-dependent gain/loss sections, on the two sides of an *imperfect* half-wave plate, which assures that rather than complete swap of polarization components there is leakage. Light is routed through this system using standard polarization beam splitters. For our experimental demonstration of this asymmetric omni-polarizer, we considered a more practical configuration, without amplifiers, i.e., $\gamma_3 = \gamma_4 = 0$, and we used optical attenuators with $\gamma_1 = \gamma_2 = 20\,\text{dB}$, and a quarter-wave plate to simulate the effect of a leaky half-wave plate with 50:50 ratio. These choices make our omni-polarizer design the direct equivalent of the asymmetric mode-switch in Fig. 2(b), but operating on orthogonal polarization states instead of guided modes. As illustrated in Fig. 3(a), to select the desired input polarization state, the angle of the input linearly-polarized light is controlled using a rotating non-leaky half-wave plate (or quarter-wave plate, as discussed in the following).

Figs. 3(b) and 3(c) report the experimental measurements of the transmission amplitude in the two output polarization states for propagation from left to right and from right to left, respectively, as a function of the input linear-polarization state, namely, the angle of the rotating half-wave plate in Fig. 3(a). For the sake of clarity, we changed the notation of the channels from "A" and "B" (used for the asymmetric mode-switch) to "V" and "H" denoting the vertical and horizontal linear polarization states. Considering Fig. 3(b) as an example, the top (bottom) solid curve is the measured transmission amplitude in the vertical (horizontal) linear-polarization channel, whereas the dashed lines are the calculated elements of the transmission matrix for all possible input-output polarization "transformations" (horizontal to vertical, vertical to vertical, horizontal to horizontal, and vertical to horizontal). The plots also show the input and output polarization ellipses. From the results in Figs. 3(b) and 3(c) it is clear that the output polarization state only depends on the propagation direction and not the input polarization: the transmission is much higher in the vertical polarization channel for propagation from left to right (Fig. 3(b)), whereas it is much higher in the horizontal polarization channel from right to left (Fig. 3(c)). The same behavior is obtained if the input state is not linearly polarized, as we experimentally demonstrated by replacing the rotating half-wave plate in Fig. 3(a) with a rotating quarter-wave plate. The transmission measurements for this case are reported in Figs. 3(d,e), which further demonstrate that the output state is independent of the input state. All these experimental results show that the proposed device indeed operates as a linear asymmetric omni-polarizer. We also stress that this design is based on an operational mechanism far simpler than any schemes relying on EP encirclements, which require complex ad-hoc structures and very long propagation lengths [28], [30].

Most importantly, our experimental results in Figs. 3(b-e) show that, although this device operates as an omni-polarizer, the transmission amplitude is *strongly dependent* on the input state, which is perfectly in line with the performance of the asymmetric mode-switch described in Section 3. Considering again Fig. 3(b) as an example, the top solid curve (vertical output) changes by more than 20 dB as the input polarization angle varies, while remaining approximately 20 dB higher than the other solid curve (horizontal output). The color of the top solid curve changes from blue to red to indicate that we go from a situation with horizontal-to-vertical polarization conversion (with high transmission) to a situation with vertical-to-vertical polarization conversion (with much lower transmission, but higher than the vertical-to-horizontal conversion). Thus, as expected, the asymmetric omni-polarizer inherits the same limitations we discussed for the asymmetric mode-switch in the previous section. We reiterate that such a large transmission matrix asymmetry is inherent to any linear reciprocal device designed to act as an asymmetric mode-switch or omni-polarizer, regardless of whether it is based on an actual EP encirclement or some other scheme, since these limitations originate directly from the reciprocity principle as discussed in Section 2.

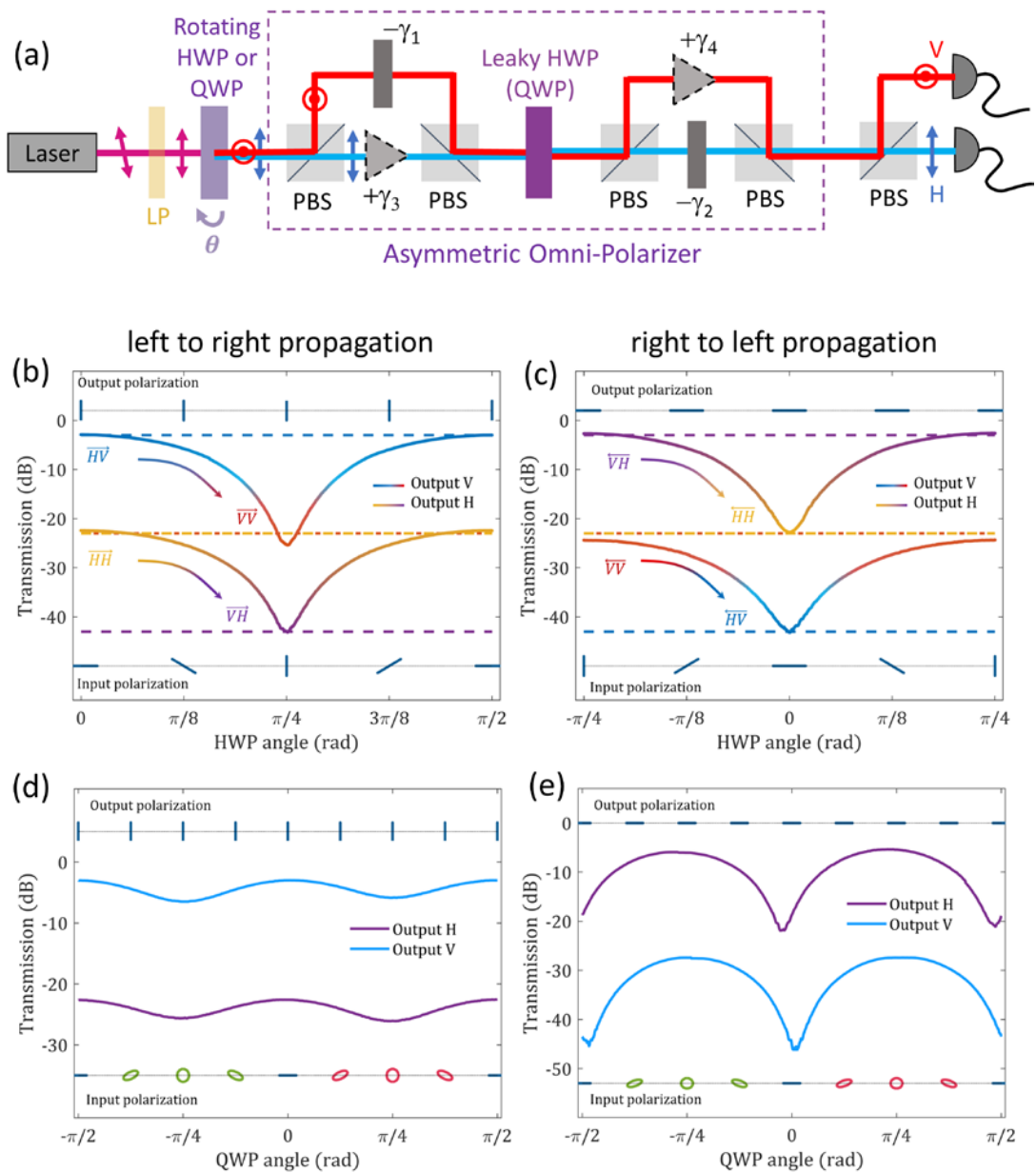

Figure 3. (a) Proposed scheme to realize an asymmetric omni-polarizer, experimentally realized with off-the-shelf optical components. (b-e) Experimental results demonstrating the operation of the proposed device as an asymmetric omni-polarizer with large transmission-matrix asymmetry. Solid lines: Experimentally measured transmission amplitude in the two output polarization states, i.e., horizontal (H)

or vertical (V) linear polarization, for propagation from left to right (b,d) and from right to left (c,e), as a function of the input linear-polarization state, namely, the angle of the rotating half-wave plate (b,c) or quarter-wave plate (d,e) in Fig. 3(a). Dashed lines in panels (b),(c): Calculated elements of the transmission matrix for all possible input-output polarization "transformations": horizontal to vertical (blue), vertical to vertical (red), horizontal to horizontal (orange), and vertical to horizontal (purple). The color gradients in the solid curves in panels (b),(c), corresponding to the colors of the dashed lines, indicate the change in polarization transformation, as described in the text. All panels also show the input and output polarization ellipses, obtained from experimental measurements.

## 5. CONCLUSION

In summary, based on theoretical considerations and experiments, we have shown that, at least in optics, the effects of a dynamic EP encirclement on the system's transmission matrix are essentially indistinguishable from the effects of a signal leakage followed by large gain/loss asymmetry. Two sets of experiments, for an asymmetric mode-switch and an asymmetric omni-polarizer, were performed using a small number of standard off-the-shelf optical components, demonstrating the same (or better) performance as actual encirclement schemes. The operation of these systems can be understood with straightforward theoretical considerations without involving the sophisticated theoretical concepts that are usually invoked when describing EP encirclements. Finally, we have shown that, due to the inherent asymmetry of the transmission response, the SNR in all the "real" and "emulated" dynamic encirclement schemes is always expected to sharply deteriorate, for at least one of two input states, which, in our view, raises serious questions about the practical applicability of EP encirclement, but perhaps does not entirely preclude some niche applications depending on the specific requirements.

**Funding.** National Science Foundation (NSF) (1741694).
**Funding.** National Science Foundation (NSF) (1741694).

**REFERENCES**

1. Bender, C.M. and S. Boettcher, *Real Spectra in Non-Hermitian Hamiltonians Having $\mathsc{P}\mathsc{T}$ Symmetry.* Physical Review Letters, 1998. **80**(24): p. 5243-5246.
2. Bender, C.M., D.C. Brody, and H.F. Jones, *Complex Extension of Quantum Mechanics.* Physical Review Letters, 2002. **89**(27): p. 270401.
3. Moiseyev, N., *Non-Hermitian quantum mechanics*. 2011, Cambridge ; New York: Cambridge University Press. xiii, 394 p.
4. Heiss, W.D., *Exceptional points of non-Hermitian operators.* Journal of Physics a-Mathematical and General, 2004. **37**(6): p. 2455-2464.
5. Ruschhaupt, A., F. Delgado, and J.G. Muga, *Physical realization of -symmetric potential scattering in a planar slab waveguide.* Journal of Physics A: Mathematical and General, 2005. **38**(9): p. L171-L176.
6. El-Ganainy, R., et al., *Theory of coupled optical PT-symmetric structures.* Optics Letters, 2007. **32**(17): p. 2632-2634.
7. Musslimani, Z.H., et al., *Optical Solitons in $\mathcal{P}\mathcal{T}$ Periodic Potentials.* Physical Review Letters, 2008. **100**(3): p. 030402.
8. Makris, K.G., et al., *Beam Dynamics in $\mathcal{P}\mathcal{T}$ Symmetric Optical Lattices.* Physical Review Letters, 2008. **100**(10): p. 103904.
9. Guo, A., et al., *Observation of $\mathcal{P}\mathcal{T}$-Symmetry Breaking in Complex Optical Potentials.* Physical Review Letters, 2009. **103**(9): p. 093902.
10. Rüter, C.E., et al., *Observation of parity–time symmetry in optics.* Nature Physics, 2010. **6**(3): p. 192-195.
11. Hodaei, H., et al., *Parity-time–symmetric microring lasers.* Science, 2014. **346**(6212): p. 975-978.
12. Feng, L., et al., *Nonreciprocal Light Propagation in a Silicon Photonic Circuit.* Science, 2011. **333**(6043): p. 729-733.
13. Feng, L., R. El-Ganainy, and L. Ge, *Non-Hermitian photonics based on parity–time symmetry.* Nature Photonics, 2017. **11**(12): p. 752-762.
14. Zyablovsky, A.A., et al., *PT-symmetry in optics.* Physics-Uspekhi, 2014. **57**(11): p. 1063-1082.
15. Wen, J., et al., *Parity-time symmetry in optical microcavity systems.* Journal of Physics B: Atomic, Molecular and Optical Physics, 2018. **51**(22): p. 222001.
16. Klaiman, S., U. Guenther, and N. Moiseyev, *Visualization of branch points in PT-symmetric waveguides.* Physical Review Letters, 2008. **101**(8).
17. Chen, W., et al., *Exceptional points enhance sensing in an optical microcavity.* Nature, 2017. **548**(7666): p. 192-196.
18. Wiersig, J., *Enhancing the Sensitivity of Frequency and Energy Splitting Detection by Using Exceptional Points: Application to Microcavity Sensors for Single-Particle Detection.* Physical Review Letters, 2014. **112**(20): p. 203901.
19. Hokmabadi, M.P., et al., *Non-Hermitian ring laser gyroscopes with enhanced Sagnac sensitivity.* Nature, 2019. **576**(7785): p. 70-74.
20. Latinne, O., et al., *Laser-Induced Degeneracies Involving Autoionizing States in Complex Atoms.* Physical Review Letters, 1995. **74**(1): p. 46-49.
21. Mailybaev, A.A., O.N. Kirillov, and A.P. Seyranian, *Geometric phase around exceptional points.* Physical Review A, 2005. **72**(1).

22. Dembowski, C., et al., *Experimental observation of the topological structure of exceptional points.* Physical Review Letters, 2001. **86**(5): p. 787-790.
23. Lee, S.B., et al., *Observation of an Exceptional Point in a Chaotic Optical Microcavity.* Physical Review Letters, 2009. **103**(13).
24. Gao, W., et al., *Continuous transition between weak and ultrastrong coupling through exceptional points in carbon nanotube microcavity exciton–polaritons.* Nature Photonics, 2018. **12**(6): p. 362-367.
25. Uzdin, R., A. Mailybaev, and N. Moiseyev, *On the observability and asymmetry of adiabatic state flips generated by exceptional points.* Journal of Physics a-Mathematical and Theoretical, 2011. **44**(43).
26. Gilary, I., A.A. Mailybaev, and N. Moiseyev, *Time-asymmetric quantum-state-exchange mechanism.* Physical Review A, 2013. **88**(1).
27. Milburn, T.J., et al., *General description of quasiadiabatic dynamical phenomena near exceptional points.* Physical Review A, 2015. **92**(5).
28. Hassan, A.U., et al., *Dynamically Encircling Exceptional Points: Exact Evolution and Polarization State Conversion.* Physical Review Letters, 2017. **118**(9): p. 093002.
29. Doppler, J., et al., *Dynamically encircling an exceptional point for asymmetric mode switching.* Nature, 2016. **537**(7618): p. 76-79.
30. Lopez-Galmiche, G., et al. *Omnipolarizer Action via Encirclement of Exceptional Points*. in *Conference on Lasers and Electro-Optics*. 2020. Washington, DC: Optical Society of America.
31. Fatome, J., et al., *A universal optical all-fiber omnipolarizer.* Scientific Reports, 2012. **2**(1): p. 938.